\documentclass[epjST]{svjour}
\usepackage{amsmath}
\usepackage{color}
\usepackage{amssymb}
\usepackage{graphicx}
\usepackage{graphics}
\begin{document}
\title{Evolution of Goldstone mode in binary condensate mixtures}
\author{Arko Roy\inst{1}~\inst{3}\fnmsep\thanks{\email{arkoroy@prl.res.in}} 
       \and S. Gautam\inst{2} 
       \and D. Angom\inst{1} }
\institute{Physical Research Laboratory, Navrangpura, Ahmedabad-380009, 
           Gujarat, India
           \and Instituto de F\'{\i}sica Te\'orica, UNESP - Universidade 
           Estadual Paulista, \\ 01.140-070 S\~ao Paulo, S\~ao Paulo, Brazil
           \and Indian Institute of Technology,
                Gandhinagar, Ahmedabad-382424, Gujarat, India}
\abstract{
We show that the third Goldstone mode in the two-species condensate mixtures, 
which emerges at phase-separation, gets hardened when the confining potentials
have separated trap centers. The {\em sandwich} type condensate density 
profiles, in this case, acquire a {\em side-by-side} density profile 
configuration. We use Hartree-Fock-Bogoliubov theory with Popov approximation 
to examine the mode evolution and density profiles for these phase 
transitions at $T=0$.
}
\maketitle
%

%%%%%%%%%%%%%%%%%%%%%%%%%%%%%%%%%%%%%%%%%%%%%%%%%%%%%%%%%%%%%%%%%%%%%%%%%%%%
%%%%                     Section: Introduction                          %%%%
%%%%%%%%%%%%%%%%%%%%%%%%%%%%%%%%%%%%%%%%%%%%%%%%%%%%%%%%%%%%%%%%%%%%%%%%%%%%

\section{Introduction}
\label{intro}

Two-species Bose-Einstein condensates (TBECs) are uniquely different from
the single-species BEC as they can be either in miscible or immiscible
phase depending upon the interaction strengths~\cite{ho_96}. 
It is, however, possible to tune a TBEC from miscible to immiscible domain or 
vice-versa in experiments \cite{tojo_10} through Feshbach resonance. 
In the last decade or so, TBECs have been realized experimentally in mixtures 
of two different alkali atoms ~\cite{mccarron_11,pasquiou_13}, or two 
different isotopes ~\cite{inouye_98} and hyperfine states
~\cite{stamper_kurn_98}. Such an unprecedented control over experiments has 
motivated numerous theoretical investigations on the stationary 
states ~\cite{gautam-10,gautam-11}, dynamical instabilities 
~\cite{sasaki_09,gautam_10}
and collective excitations ~\cite{stringari_96,ticknor_14} of TBECs.

The present work describes the development of a gapless~\cite{hugenholtz_59} 
Hartree-Fock-Bogoliubov theory with Popov (HFB-Popov) approximation
\cite{griffin_96} for TBECs. We apply it to examine the spectra of the 
quasi-particle excitations of trapped TBEC in separated and non-separated trap 
centers. In the first part of the work we study the hardening of the 
Goldstone mode with the increase in the separation of the trap centers in
which the TBECs have been confined. This has profound experimental 
implications since, in experiments, the trap centers never coincide. This is
due to gravitational sagging, and deviations of the trapping potentials
from perfect alignment. In the later part of our study, we show the softening 
of Goldstone mode with the variation in inter-species interaction strengths, 
albeit, with co-incident trap centers. Previous studies 
have shown the existence of an additional Goldstone mode in the excitation 
spectrum at phase-separation for the transition from miscible to 
symmetry-broken or {\em side-by-side} density profiles of TBEC 
~\cite{ticknor_14,ticknor_13,takeuchi_13}. We, here, show that in the
the changeover from miscible to {\em sandwich} type density profile, where one 
of the species is enveloped by the other, is accompanied by the appearance of a
third Goldstone mode at the transition point. Even at higher inter-species
interactions, the mode continues to be a Goldstone mode.

%%%%%%%%%%%%%%%%%%%%%%%%%%%%%%%i%%%%%%%%%%%%%%%%%%%%%%%%%%%%%%%%%%%%%%%%%%%%%%%%
%%%%%%%%%%%                    Theory                       %%%%%%%%%%%%%%%%%%%
%%%%%%%%%%%%%%%%%%%%%%%%%%%%%%%%%%%%%%%%%%%%%%%%%%%%%%%%%%%%%%%%%%%%%%%%%%%%%%%

\section{Theory of two component BEC}
\label{theory}
To provide a coherent description of the HFB-Popov theory for a quasi-1D TBEC, 
we start with the theory for a single-species BEC. In a cigar-shaped harmonic 
trap for which the radial frequency
$\omega_{\perp} (\omega_x = \omega_y = \omega_{\perp}) 
 \gg$ the axial frequency $\omega_z$, the time-independent 1D generalized GP
equation with the HFB-Popov approximation is given by \cite{griffin_96}
\begin{equation}
  \hat{h}\phi + U\left[n_{c}+2\tilde{n}\right]\phi = 0,
  \label{gpe1s}
\end{equation}
where $\hat{h}= (-\hbar^{2}/2m)\partial ^2/\partial z^2 + V(z)-\mu$. The
strength of the repulsive contact interaction is represented by 
$U = (a\lambda)/m$ with $a$ as the $s$-wave scattering length, and  
$\lambda = (\omega_{\perp}/\omega_z) \gg 1$ quantifies the anisotropy of
the trapping potential. The atomic mass of the species is given by $m$, 
and $\mu$ is the chemical potential. The quantities
$n_{c}(z)\equiv|\phi(z)|^2$,
$\tilde{n}(z)\equiv\langle\tilde{\psi}^{\dagger}(z,t)
\tilde{\psi}(z,t)\rangle$, and $n(z) = n_{c}(z)+ \tilde{n}(z)$ in the above
equation represent the  condensate, non-condensate, and total density,
respectively along the axial direction of the trap. In the equation
$\phi(z)$ is the order parameter and $\tilde\psi(z)$ is the field operator
associated with the fluctuation. The non-condensate or thermal component 
$\tilde{n}(z)$ is computed from the solutions of the Bogoliubov-de-Gennes 
equations
\begin{eqnarray}
\nonumber
(\hat{h}+2Un)u_{j}-U\phi^{2}v_{j}&=&E_{j}u_{j},\\
-(\hat{h}+2Un)v_{j}+U\phi^{*2}u_{j}&=& E_{j}v_{j}.
\label{bdg1}
\end{eqnarray}
Here $u_j$'s and $v_j$'s are the Bogoliubov quasi-particle amplitudes, 
and $E_j$s are the eigen energy of the $j$th quasi-particle excitation.
Applying the same formalism, the ground state of a quasi-1D TBEC
is described by a pair of coupled generalized GP equations~\cite{roy_14},
\begin{eqnarray}
 \hat{h}_k\phi_k + U_{kk}\left(n_{ck}+2\tilde{n}_{k}\right)\phi_k
  +U_{12}n_{3-k}\phi_k=0.
\label{gpe}
\end{eqnarray}
Here $\hat{h}_{k}= (-\hbar^{2}/2m_k)\partial ^2/\partial z^2 + V_k(z)-\mu_k$
and $k=1,2$ is the species index. The quantities 
$U_{kk} = (a_{kk}\lambda)/m_{k}$ and $U_{12}=(a_{12}\lambda)/(2m_{12})$, are
the intra- and inter-species interactions, respectively, with
$m_{12}=m_1 m_2/(m_1+m_2)$ as the reduced mass. For the present study, we
consider repulsive interactions, that is $a_{kk},a_{12} > 0$. The $\phi_k$s 
are the order parameters with $n_{ck}(z)\equiv|\phi_k(z)|^2$,
$\tilde{n}_k(z)\equiv\langle\tilde{\psi}_{k}^{\dagger}(z,t)
\tilde{\psi}_k(z,t)\rangle$, and $n_k(z) = n_{ck}(z)+ \tilde{n}_k(z)$
as the local condensate, non-condensate, and total density, respectively of
the $k$th species. Like in single species case, let $\tilde\psi_{k}(z)$s 
represent the fluctuations, then, the non-condensate density $\tilde{n}_k(z)$ 
is obtained from the solutions of the coupled Bogoliubov-de-Gennes equations
%\begin{subequations}
\begin{eqnarray}
\nonumber
 \hat{{\mathcal L}}_{1}u_{1j}-U_{11}\phi_{1}^{2}v_{1j}+U_{12}\phi_1 \left 
  (\phi_2^{*}u_{2j} -\phi_2v_{2j}\right )&=& E_{j}u_{1j},\;\;\;\;\;\;\\
  \nonumber
 \hat{\underline{\mathcal L}}_{1}v_{1j}+U_{11}\phi_{1}^{*2}u_{1j}-U_{12}\phi_1^*\left (
   \phi_2v_{2j}-\phi_2^*u_{2j} \right ) &=& E_{j}v_{1j},\;\;\;\;\;\;\\
   \nonumber
 \hat{{\mathcal L}}_{2}u_{2j}-U_{22}\phi_{2}^{2}v_{2j}+U_{12}\phi_2\left ( 
   \phi_1^*u_{1j}-\phi_1v_{1j} \right ) &=& E_{j}u_{2j},\;\;\;\;\;\;\\
\label{bdg}
 \hat{\underline{\mathcal L}}_{2}v_{2j}+U_{22}\phi_{2}^{*2}u_{2j}-U_{12} \phi_2^*\left ( 
 \phi_1v_{1j}-\phi_1^*u_{1j}\right ) &=& E_{j}v_{2j},\;\;\;\;\;\; 
\end{eqnarray}
%\end{subequations}
where $\hat{{\mathcal L}}_{1}=
\big(\hat{h}_1+2U_{11}n_{1}+U_{12}n_{2})$, $\hat{{\mathcal L}}_{2}
=\big(\hat{h}_2+2U_{22}n_{2}+U_{12}n_{1}\big)$ and
$\hat{\underline{\cal L}}_k  = -\hat{\cal L}_k$. Here $u_k$s and $v_k$s are 
the Bogoliubov quasi-particle amplitudes, and $j$ is the energy eigenvalue 
index. Taking the quantum fluctuations into account, the order parameters 
$\phi_k$s are 
calculated self-consistently using Eqns.~\ref{gpe},~\ref{bdg} as elaborated in 
Ref.~\cite{roy_14}. In the present study, we examine the ground state of the 
TBEC and compute the energy eigenmodes at zero temperature both in miscible
($U_{12}<\sqrt{U_{11}U_{22}}$) and immiscible ($U_{12}>\sqrt{U_{11}U_{22}}$)
regimes with co-incident and non co-incident trap centers. 

\begin{figure}
\begin{center}
\resizebox{1.0\hsize}{!}
 {\includegraphics{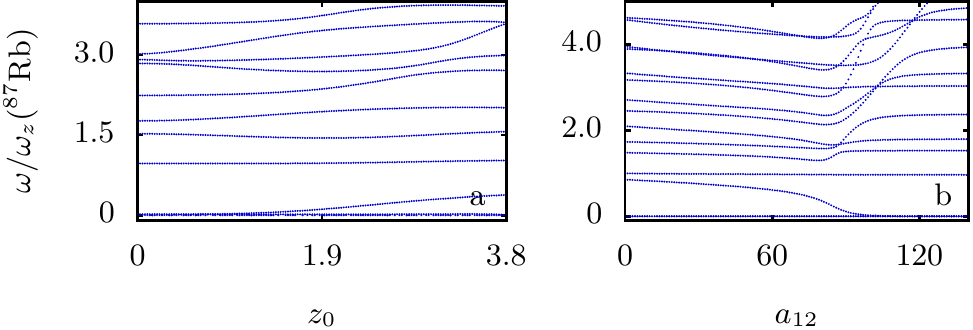}}%{mode_evol_delta_f}
 \caption{The evolution of the low-lying quasiparticle  eigen frequencies 
          in the Rb-Na TBEC; $N_{^{23}{\rm Na}} = N_{^{87}{\rm Rb}} = 10^4$.
          (a) Shows the evolution of the low-lying quasiparticle excitations 
          as a function of $z_0$, trap center separation, in
          the domain $0\leqslant z_0 \leqslant 3.8 a_{\rm osc (Rb)} $ for
          $a_{_{\rm NaRb}}=100 a_0$, 
          (b) Shows the evolution of the low-lying quasiparticle  
              excitations as a function of $a_{\rm NaRb}$ in
          the domain $0\leqslant a_{\rm NaRb}\leqslant 150 a_0 $ for
          $z_0=0$.
          }
\label{mode_evol_delta}
\end{center}
\end{figure}

%%%%%%%%%%%%%%%%%%%%%%%%%%%%%%%%%%%%%%%%%%%%%%%%%%%%%%%%%%%%%%%%%%%%%%%%%%%%%%%
%%%%            Results and Discussions                                  %%%%%% 
%%%%%%%%%%%%%%%%%%%%%%%%%%%%%%%%%%%%%%%%%%%%%%%%%%%%%%%%%%%%%%%%%%%%%%%%%%%%%%%

\section{Results and Discussions}

%%%%%%%%%%%%%%%%%%%%%%%%%%%%%%%%%%%%%%%%%%%%%%%%%%%%%%%%%%%%%%%%%%%%%%%%%%%%%%%
%%%%   Mode hardening in $^{87}$Rb-$^{23}$Na TBEC at $T=0$ due to           %%% 
%%%%              displaced trap centers                                    %%%
%%%%%%%%%%%%%%%%%%%%%%%%%%%%%%%%%%%%%%%%%%%%%%%%%%%%%%%%%%%%%%%%%%%%%%%%%%%%%%%

\subsection{Mode hardening in $^{87}$Rb-$^{23}$Na TBEC at $T=0$ due to 
            displaced trap centers}

In experiments, the gravitational potential of Earth and tilts in the
external trapping potentials displace the minima of the traps. The
potentials are no longer co-incident and can be replaced by the following
effective potentials 
\begin{equation}
 V_k(z,z_0) = 1/2m_k\omega_{zk}^2\left [ z+(-1)^kz_0\right ] ^2,
\end{equation}
where $2z_0$ is the separation between two trap centers, and the other
symbols have their usual meaning. It is, therefore, pertinent to examine the
variation of energies of the quasi-particle excitations and the topological
deformation of the density profiles with the change in
the separation of the trap centers. For our present study, we consider a
quasi-1D Rb-Na mixture in the immiscible regime 
with $\omega_{z({\rm Rb})} = 2\pi\times 4.55 $Hz,
$\omega_{z({\rm Na})} = 2\pi\times 3.89 $Hz and
$\omega_{\perp({\rm Rb})} = 2\pi\times 40.2 $Hz,
$\omega_{\perp({\rm Na})} = 2\pi\times 32.2 $Hz
and $N_{\rm Na} = N_{\rm Rb}=10^4$. Let Rb and Na be the first and
second species, respectively with $a_{11}=a_{\rm RbRb} = 100 a_0$,
$a_{22}= a_{\rm NaNa}=50a_0$, and $a_{12}=a_{\rm NaRb} = 100 a_0$ 
where $a_0$ is the Bohr radius. When the trap centers are co-incident, that is,
$z_0=0$, the spectra is characterized by three Goldstone modes. The condensate
density profile assumes a sandwich geometry, in which Rb condensate is at the 
center and flanked by Na condensate at the edges. 

\begin{figure}
\begin{center}
\resizebox{0.7\hsize}{!}
 {\includegraphics{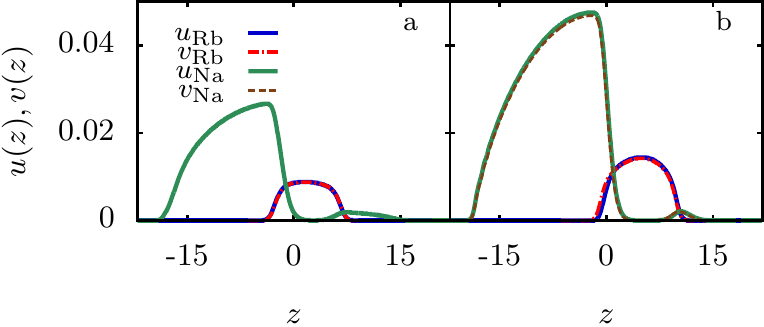}}
 \caption{The evolution of the low-lying quasiparticle  eigen functions as
          a function of $z_0$ in
          the Rb-Na TBEC in the domain 
          $0\leqslant z_0 \leqslant 3.8 a_{\rm osc (Rb)} $ for
          (a) $z_0=1.5a_{\rm osc (Rb)}$, (b) $z_0=2.5a_{\rm osc (Rb)}$.
         }
\label{efunction}
\end{center}
\end{figure}

For $z_0>0$, separated trap centers, with the breaking of the $z$-parity of 
the system, the energy of the second Goldstone mode of the Na
condensate gradually increases and gets hardened at a critical value of
$z_0$. The Na Kohn mode which is transformed into the  third Goldstone 
mode of the system, second for the Na condensate, at phase separation
with $z=0$, regains energy  when $z_0>0$ to emerge as the second Kohn mode of 
the system. 
This is evident from the mode evolution as shown as a function of
$z_0$ in Figs.~\ref{mode_evol_delta}(a), and the profile of the hardened
mode mode is shown in Fig. \ref{efunction}.
The energy of the Rb Kohn mode, on the other hand, remains unchanged even in a
separated trap setting. Furthermore, with the change from co-incident to
non con-incident trap center, the sandwich type density profile 
gradually changes into a side-by-side density profile. The excitation spectra
of the system is now identified by two Goldstone modes and two Kohn modes.
It is to be recalled here that, throughout the transition, the TBECs are 
always phase separated. It must, however, be emphasized that the evolution of 
the Goldstone mode when the TBEC undergoes a transition from miscible to 
immiscible with $z_0=0$ is different from the current case. We reported this
in our previous work \cite{roy_14} for Rb-Cs TBEC, and in the present work
we discuss the same for the Rb-Na in the following section.

%%%%%%%%%%%%%%%%%%%%%%%%%%%%%%%%%%%%%%%%%%%%%%%%%%%%%%%%%%%%%%%%%%%%%%%%%%%%%%%
%%%%     Third Goldstone mode in $^{87}$Rb-$^{23}$Na TBEC at $T=0$       %%%%%% 
%%%%%%%%%%%%%%%%%%%%%%%%%%%%%%%%%%%%%%%%%%%%%%%%%%%%%%%%%%%%%%%%%%%%%%%%%%%%%%%

\subsection{Third Goldstone mode in $^{87}$Rb-$^{23}$Na TBEC at $T=0$}

The transition from miscible to immiscible phases in TBECs occurs when
$U_{12}> \sqrt{U_{11}U_{22}}$. For the Rb-Na TBEC, experimentally, it is 
possible to steer the system from miscible to immiscible domain through the 
Rb-Na Feshbach resonance~\cite{wang_13}. This motivates us to examine the 
variation of eigenenergies $E_j$s of the quasi-particle excitations at, and 
around the point of phase separation. We thus vary $a_{_{\rm NaRb}}$ in the 
domain $0\leqslant a_{_{\rm NaRb}}\leqslant150 a_0$ to study the role of 
inter-species interactions in the Rb-Na TBEC, more precisely, the effect on 
the excitation spectra. In the absence of inter-species interactions, that is 
$a_{_{\rm NaRb}}=0$, the excitation spectrum has two Goldstone modes, and 
two Kohn modes with eigen energies $ \hbar\omega_{z({\rm Rb})}$ and 
$0.85 \hbar\omega_{z({\rm Rb})}$ for Rb and Na species, respectively.
Further more, the modes of the two species are distinct and uncoupled. 

  For $a_{_{\rm NaRb}} > 0$, the eigen spectra of the two species are not
independent, and as a result the modes intermix. The energy of the Rb Kohn 
mode remains unchanged with the variation in $a_{_{\rm NaRb}}$. The energy of 
the Na Kohn mode, however, decreases with the increase of $a_{_{\rm NaRb}}$, 
and goes soft at a critical value of $a_{_{\rm NaRb}}$ when the 
phase-separation occurs. This generates a new Goldstone mode of the Na BEC in 
the Bogoliubov excitation spectrum as shown in Fig.\ref{mode_evol_delta}(b). 
This happens because at phase-separation, the miscible density profiles 
transform to {\em sandwich} type density profiles, with the $^{87}$Rb 
condensate located at the center and $^{23}$Na condensate at the edges. Thus, 
effectively the Na condensate consist of two topologically disconnected BECs 
and there are two Goldstone modes with the same $|u_{\rm Na}|$ and 
$|v_{\rm Na}|$, but different phases~\cite{roy_14}. For the other case,
transition from miscible to {\em side-by -side} density profiles, the Kohn 
mode energy of one of the species decreases, and goes soft at phase-separation.
This breaks the $z$-parity symmetry of the system, and the zero-energy mode 
then starts to regain energy after phase-separation \cite{ticknor_13}.
Detailed discussions on Goldstone modes and bifurcations in TBECs with 
or without soliton are discussed in our previous works\cite{roy_14,roy_14a}.

%%%%%%%%%%%%%%%%%%%%%%%%%%%%%%%%%%%%%%%%%%%%%%%%%%%%%%%%%%%%%%%%%%%%%%%%%%%%%%%
%%%%                              Conclusions                           %%%%%% 
%%%%%%%%%%%%%%%%%%%%%%%%%%%%%%%%%%%%%%%%%%%%%%%%%%%%%%%%%%%%%%%%%%%%%%%%%%%%%%%

\section{Conclusions}
The third  Goldstone mode in TBECs with {\em sandwich} profile at phase 
separation gets harden when the trap centers are separated by a critical 
distance. This is accompanied by the topological change from {\em sandwich} to
{\em side-by-side} condensate density profiles. This result has important 
experimental implications, and demonstrates why it is a major challenge to 
obtain {\em sandwich} type density profiles in TBEC experiments. The 
appearance of a third Goldstone mode in TBECs with {\em sandwich} type density 
profile at phase-separation in Rb-Na is another finding, which confirms and 
adds to the similar result reported in our previous work \cite{roy_14}. This 
Goldstone mode is associated with the fragmentation of the condensate species 
at the edges, Na in the present work, into two distinct condensates. So, 
effectively, the TBEC at phase separation with {\em sandwich } geometry, in 
quasi-1D, is equivalent to three coupled condensate clouds, and thus three 
Goldstone modes emerge in the excitation spectrum.

%%%%%%%%%%%%%%%%%%%%%%%%%%%%%%%%%%%%%%%%%%%%%%%%%%%%%%%%%%%%%%%%%%%%%%%%%%%%%%%
%%%%                           Acknowledgement                           %%%%%% 
%%%%%%%%%%%%%%%%%%%%%%%%%%%%%%%%%%%%%%%%%%%%%%%%%%%%%%%%%%%%%%%%%%%%%%%%%%%%%%%
\section*{Acknowledgements}
We thank K. Suthar and S. Chattopadhyay for useful
discussions. The results presented in the paper are based on the
computations using the 3TFLOP HPC Cluster at Physical Research Laboratory,
Ahmedabad, India.

%%%%%%%%%%%%%%%%%%%%%%%%%%%%%%%%%%%%%%%%%%%%%%%%%%%%%%%%%%%%%%%%%%%%%%%%%%%%%%%
%%%%                           References                               %%%%%% 
%%%%%%%%%%%%%%%%%%%%%%%%%%%%%%%%%%%%%%%%%%%%%%%%%%%%%%%%%%%%%%%%%%%%%%%%%%%%%%%

\end{document}